\documentclass[sigconf,screen]{acmart}

\AtBeginDocument{%
  \providecommand\BibTeX{{%
    \normalfont B\kern-0.5em{\scshape i\kern-0.25em b}\kern-0.8em\TeX}}}

\usepackage{microtype}
\usepackage{balance}
\usepackage{enumitem}
\usepackage{soul}
\usepackage{xspace}

\usepackage{graphicx}
\usepackage{xcolor}
\usepackage{colortbl}
\usepackage{array}
\usepackage{booktabs}
\usepackage{diagbox}
\usepackage{makecell}
\usepackage{multirow}
\usepackage{tabularx}
\usepackage[flushleft]{threeparttable}
\usepackage{subcaption}
\usepackage{adjustbox}
\usepackage{tcolorbox}
\usepackage{mdframed}

\usepackage{listings}

\definecolor{pblue}{rgb}{0.13,0.13,1}
\definecolor{pgreen}{rgb}{0,0.5,0}
\definecolor{pred}{rgb}{0.9,0,0}
\definecolor{pgrey}{rgb}{0.46,0.45,0.48}
\definecolor{applegreen}{rgb}{0,0.5,0.0}

\definecolor{codeblue}{RGB}{20,76,134}
\definecolor{codegreysh}{RGB}{114,136,223}
\definecolor{code}{RGB}{51,51,255}

\definecolor{prefixblue}{RGB}{46,117,182}
\definecolor{prefixbluebg}{RGB}{219,233,246}
\definecolor{mutgreen}{RGB}{39,174,96}
\definecolor{mutgreenbg}{RGB}{212,239,223}
\definecolor{docpurple}{HTML}{6A1B9A}
\definecolor{docpurplebg}{RGB}{235,218,242}

\definecolor{togablue}{HTML}{1F4E79}
\definecolor{togllgreen}{HTML}{2E7D32}
\definecolor{rqblue}{HTML}{1F4E79}
\definecolor{rqgreen}{HTML}{2E7D32}
\definecolor{rqpurple}{HTML}{6A1B9A}

\lstdefinestyle{listingstyle}{
  language=Java,
  basicstyle=\ttfamily\scriptsize,
  keywordstyle=\bf\ttfamily\color{codeblue},
  stringstyle=\color{codegreysh},
  moredelim=[l][\bf\ttfamily\color{red}]{///},
  moredelim=[l][\bf\ttfamily\color{orange}]{//,},
  moredelim=[s][\bf\ttfamily\color{code}]{/**}{**/}
}

\newcommand{\stepbox}[2]{%
  \fcolorbox{black!25}{#1}{\strut\footnotesize #2}%
}

\providecommand{\modelcell}{}
\renewcommand{\modelcell}[3]{%
  \makecell[c]{\textbf{\textcolor{#1}{\textsc{#2}}}\\[-1pt]
  {\scriptsize\textbf{\textcolor{#1}{#3}}}}%
}

\newcommand{\rqmodelcell}[3]{%
  \makecell[c]{\textbf{\textcolor{#1}{\textsc{#2}}}\\[-1pt]
  {\scriptsize\textbf{\textcolor{#1}{#3}}}}%
}

\setcopyright{cc}
\setcctype{by}
\acmDOI{10.1145/3832783.3837464}
\acmYear{2026}
\copyrightyear{2026}
\acmISBN{979-8-4007-2882-2/2026/10}
\acmConference[ASE '26]{Proceedings of the 41st IEEE/ACM International Conference on Automated Software Engineering}{October 12--16, 2026}{Munich, Germany}
\acmBooktitle{Proceedings of the 41st IEEE/ACM International Conference on Automated Software Engineering (ASE '26), October 12--16, 2026, Munich, Germany}
\acmSubmissionID{ase26main-p773-p}
\received{2026-03-26}
\received[accepted]{2026-06-18}

\begin{document}

\title{Documentation vs. Code Patterns: What Drives LLM-Based Exception Oracle Generation?}

\author{Soneya Binta Hossain}
\correspondingauthor
\orcid{0000-0002-7282-061X}
\affiliation{%
  \institution{University of Texas at Dallas}
  \city{Richardson}
  \country{USA}
}
\email{sbhossain@utdallas.edu}

\author{Matthew B. Dwyer}
\orcid{0000-0002-1937-1544}
\affiliation{%
  \institution{University of Virginia}
  \city{Charlottesville}
  \country{USA}
}
\email{matthewbdwyer@virginia.edu}

\author{Tasfia Tasnim}
\orcid{0009-0008-0200-7675}
\affiliation{%
  \institution{University of Texas at Dallas}
  \city{Richardson}
  \country{USA}
}
\email{tasfia.tasnim@utdallas.edu}

\newcommand{\todoc}[2]{{\textcolor{#1}{#2}}}
\newcommand{\todoblack}[1]{{\todoc{black}{\textbf{[[#1]]}}}}
\newcommand{\todored}[1]{{\todoc{red}{\textbf{[[#1]]}}}}
\newcommand{\todogreen}[1]{\todoc{green}{\textbf{[#1]}}}
\newcommand{\todoblue}[1]{\todoc{black}{#1}}
\newcommand{\todoorange}[1]{\todoc{orange}{\textbf{[[#1]]}}}
\newcommand{\todobrown}[1]{\todoc{brown}{\textbf{[[#1]]}}}
\newcommand{\todogray}[1]{\todoc{gray}{\textbf{[[#1]]}}}
\newcommand{\todopurple}[1]{\todoc{purple}{\textbf{[#1]}}}
\newcommand{\todopink}[1]{\todoc{magenta}{\textbf{[[#1]]}}}
\newcommand{\todocyan}[1]{\todoc{cyan}{\textbf{[[#1]]}}}
\newcommand{\todoviolet}[1]{\todoc{violet}{\textbf{[[#1]]}}}
\newcommand{\todo}[1]{\todored{TODO: #1}}

\newcommand{\rt}[1]{\todopink{Raygan: #1}}
\newcommand{\soneya}[1]{\todoblue{Soneya: #1}}
\newcommand{\ignore}[1]{}
\newcommand{\raygan}[1]{\todoblue{#1}}
\newcommand{\toolname}{Toggle}
\newcommand{\matt}[1]{\todopurple{Matt: #1}}
\newcommand{\mr}[1]{{\color{black} #1}}
\newcommand{\mrnew}[1]{{\color{black} #1}}

\definecolor{col1}{RGB}{240, 240, 240} 
\definecolor{col2}{RGB}{240, 240, 240} 
\definecolor{col3}{RGB}{220, 220, 220} 
\definecolor{col4}{RGB}{220, 220, 220} 
\definecolor{col5}{RGB}{200, 200, 200} 
\definecolor{col6}{RGB}{200, 200, 200} 



\newcommand{\code}[1]{\texttt{\small #1}} 

\newcounter{finding}
\newcommand{\finding}[1]{\refstepcounter{finding}
    \begin{mdframed}[linecolor=gray,roundcorner=12pt,backgroundcolor=gray!15,linewidth=3pt,innerleftmargin=2pt, leftmargin=0cm,rightmargin=0cm,topline=false,bottomline=false,rightline=false]
        \textbf{Finding \arabic{finding}:} #1
    \end{mdframed}
}

\setlist[itemize]{leftmargin=20pt}

\begin{abstract}

LLM-based test oracle generation (TOG) methods report high accuracy on exception oracle generation, but it remains unclear what evidence drives these predictions. In particular, do models use explicit exceptional-behavior documentation such as Javadoc \texttt{@throws} clauses, or do they rely on recurring patterns in tests, code, and documentation?

We investigate this question through a large-scale intervention-based study of three TOG systems spanning classifier-based and generative architectures and model sizes from roughly 110M to 7B parameters, evaluated on three real-world benchmarks comprising two generated-test datasets and a new benchmark of developer-written tests. We first remove Javadoc \texttt{@throws} clauses and find that accuracy changes only marginally, with the largest drop below one percentage point. This indicates that structured exception documentation is not the primary driver of exception-oracle prediction. We then apply attribution-guided substitution ablations to identify the signals that predictions depend on. The results show that high accuracy can be driven by shortcut signals: some models are highly sensitive to a small number of structural tokens, while others distribute reliance across many lexical cues.

These findings challenge the assumption that strong exception-oracle accuracy reflects robust use of exception semantics. Future TOG systems should therefore be evaluated not only by whether they predict the correct oracle type, but also by whether their predictions are grounded in meaningful exception-triggering evidence.

\end{abstract}

\ccsdesc[500]{Software and its engineering~Software testing and debugging}

\keywords{automated software testing, test oracle generation, exception oracles,
large language models, Javadoc, model attribution}

\maketitle

\section{Introduction}

Exceptions are fundamental in languages such as Java, C\#, Python, and C++, where they signal abnormal conditions, enforce API contracts, and help preserve correct execution. Because software must behave correctly in both normal and exceptional executions, specifying and testing exceptional behavior is critical to reliability~\cite{cabral2007exception,robillard2003static}. Yet exceptional behavior is often harder to observe than value-returning behavior~\cite{nguyen2019empirical,fu2004testing}: return values appear in ordinary executions, whereas exceptions arise only under specific runtime conditions. Developers and testing tools therefore often rely on external artifacts, especially documentation, to understand \textit{when} and \textit{why} exceptions should occur. In Java, Javadoc \texttt{@throws} clauses, including the synonymous \texttt{@exception} tag, provide one of the most structured forms of exception-related documentation. Similar mechanisms appear in other ecosystems, such as XML documentation in C\#, Python docstrings, and contract-based specifications, making exceptional-behavior documentation a broader software-engineering concern.

This makes testing exceptional behavior a central part of software validation~\cite{6963470}. A system should not only return correct values in normal executions, but also detect, signal, and propagate errors in abnormal ones. This requirement is captured through \emph{exception oracles}: test-oracle specifications that state whether a test should expect an exception. In Java, such oracles are commonly encoded using JUnit idioms such as \textsc{@Test(expected=...)} in JUnit 4, \textsc{assertThrows} in JUnit 5, or explicit \textsc{try--catch} blocks, as shown in Listing~\ref{lst:exception-oracles}~\cite{junit4test,junit5}. Similar constructs exist in other ecosystems, such as Python's \textsc{assertRaises} and NUnit's \textsc{Assert.Throws}~\cite{pyunittest,nunitdocs}. Despite their importance, exception oracles remain labor-intensive to write, error-prone, and dependent on developers anticipating failure modes that may be only partially documented~\cite{cabral2007exception}.

Documentation can help bridge this gap. Javadoc often states the conditions under which exceptions should occur, making it a natural source for constructing exception oracles. For example, in Listing~\ref{lst:javadoc} (Javadoc 1), the documentation for \textsc{ListOrderedMap.remove} states that an \textsc{IndexOutOfBoundsException} should be thrown when the index is invalid. This directly suggests tests that pass invalid indices, such as $-1$ or an index equal to the map size, and expect an \textsc{IndexOutOfBoundsException}. Listing~\ref{lst:javadoc} (Javadoc 2) illustrates the same idea for a null-triggered exception. More generally, when a Javadoc clause identifies the condition that triggers an exception, it provides actionable specification evidence that can be translated into concrete inputs and expected exceptional behavior. In this sense, Javadoc can support automated exception-oracle generation rather than merely describe exceptional behavior.

\begin{lstlisting}[float=tp,language=Java, basicstyle=\ttfamily\footnotesize, label={lst:exception-oracles}, caption={Different ways of writing exception oracles in Java.}]
// (1) Using the JUnit 4 @Test annotation
@Test(expected = IOException.class)
public void test0() throws IOException {
    MyFileReader r = new MyFileReader("none.txt");
    r.open();
}

// (2) Using JUnit 5's assertThrows
@Test
public void test1() {
    assertThrows(IOException.class, () -> {
        MyFileReader r = new MyFileReader("none.txt");
        r.open();
    });
}

// (3) Using explicit try-catch with a fail()
@Test
public void test2() {
    try {
        MyFileReader r = new MyFileReader("none.txt");
        r.open();
        fail("Expected IOException not thrown");
    } catch (IOException e) {
        // test passes if exception is thrown
    }
}
\end{lstlisting}

Recent TOG work has advanced from assertion-oriented learning approaches~\cite{watson2020learning}, to documentation-driven techniques for exceptional behavior~\cite{tan2012tcomment,goffi2016automatic}, to neural and LLM-based systems such as \textsc{TOGA}, ChatAssert, \textsc{TOGLL}, and \textsc{Doc2OracLL}~\cite{Dinella2022TOGA,hayet2024chatassert,togll-cse-25,Doc2OracLL-fse-25}. However, the basis of reported exception-oracle performance remains unclear. Prior TOG work has mainly focused on assertion oracles or on treating documentation as a broad context source, leaving the drivers of exception-oracle prediction underspecified. This matters because high accuracy alone does not show whether a model used meaningful exception evidence, such as a documented precondition, or relied on recurring shortcut cues in code, documentation, or test prefixes.

\begin{lstlisting}[float=tp, language=Java, basicstyle=\ttfamily\footnotesize, label={lst:javadoc}, caption={Examples of Javadoc exception clauses and corresponding exception-oracle tests.}]
                Javadoc (1)
/**
 * Removes the element at the specified index.
 * @param index  the index of the object to remove
 * @return the removed value, or {@code null} if none existed
 * @throws IndexOutOfBoundsException if the index is invalid 
 */
public void test0() {
    resetEmpty();
    resetFull();
    final ListOrderedMap<K,V> lom = getMap();
    assertThrows(IndexOutOfBoundsException.class, () -> lom.remove(-1));
}
                Javadoc (2)
/**
 * Creates a parameterized type instance.
 * @param rawClass the raw class to create a parameterized type instance for
 * @param typeArguments the types used for parameterization
 * @return {@link ParameterizedType}
 * @throws NullPointerException if {@code rawClass} is {@code null}
 * @since 3.2 
 */
public void test1() {
    final Class<?> rawClass = null;
    final Map<TypeVariable<?>, Type> tvMap = Collections.emptyMap();
    assertThrows(NullPointerException.class, () -> TypeUtils.parameterize(rawClass, tvMap));
}
\end{lstlisting}

In this paper, we investigate what actually drives exception-oracle prediction in neural and LLM-based TOG. We answer this through a large-scale intervention-based study of three representative TOG systems on three real-world Java datasets: two generated-test benchmarks (OE25 and Sf110) and OE25\textsubscript{dev}, a newly curated benchmark of developer-written tests from 25 systems. We define \emph{exception-oracle accuracy} as an oracle-type prediction metric: for an exception-labeled instance, a prediction is correct when the instance is classified as \textsc{Exception}, not \textsc{Assertion}. This metric does not measure semantic equivalence, compilability, oracle style, or exact exception-type matching. Section~\ref{sec:metrics} gives the formal definition.

To identify the signals behind these predictions, we compare model behavior before and after removing Javadoc \texttt{@throws} clauses, and then apply attribution-guided substitution ablations with cue categorization across the test prefix, focal code, and documentation. Removing these exception clauses (ECs) causes no change or only very small drops in accuracy: the largest observed overall drop is 0.16 percentage points, and the largest drop among samples that originally contained exceptional-behavior clauses is 0.54 percentage points. This suggests that structured exception documentation is not the main driver. Instead, attribution-guided ablation shows that models rely on alternative input signals. For \textsc{TOGLL}, correct predictions often flip after only 3--4 high-attribution substitutions and are dominated by structural cues such as newlines. For \textsc{Doc2OracLL}, predictions are harder to flip, but still depend on distributed lexical cues rather than on \texttt{@throws}/\texttt{@exception} clauses alone.

Overall, this work argues that LLM-assisted task automation should be evaluated beyond accuracy alone. For TOG, this means checking whether exception predictions are grounded in meaningful evidence---such as the triggering precondition, method under test, documentation, or feasible throw path. It also means using controlled contexts, diverse tests, hard negatives, and counterfactual cases to expose shortcut cues such as formatting patterns or recurring lexical tokens.

\textbf{In summary, this paper makes the following contributions:}
\begin{itemize}
    \item We present a large-scale intervention-based study of the signals that drive exception-oracle prediction in neural and LLM-based TOG.
    \item We introduce OE25\textsubscript{dev}, a benchmark of developer-written Java tests that complements prior generated-test datasets with more realistic test structures and oracle patterns.
    
    \item We show that Javadoc \texttt{@throws}/\texttt{@exception} clauses have limited effect, while alternative input cues play a much larger role in exception-oracle prediction.
    \item We derive practical guidance for future TOG systems: evaluate beyond accuracy, control for shortcut cues, and require predictions to be grounded in exception-triggering evidence.
\end{itemize}

\begin{figure*}[h]
    \centering
    \includegraphics[width=\linewidth]{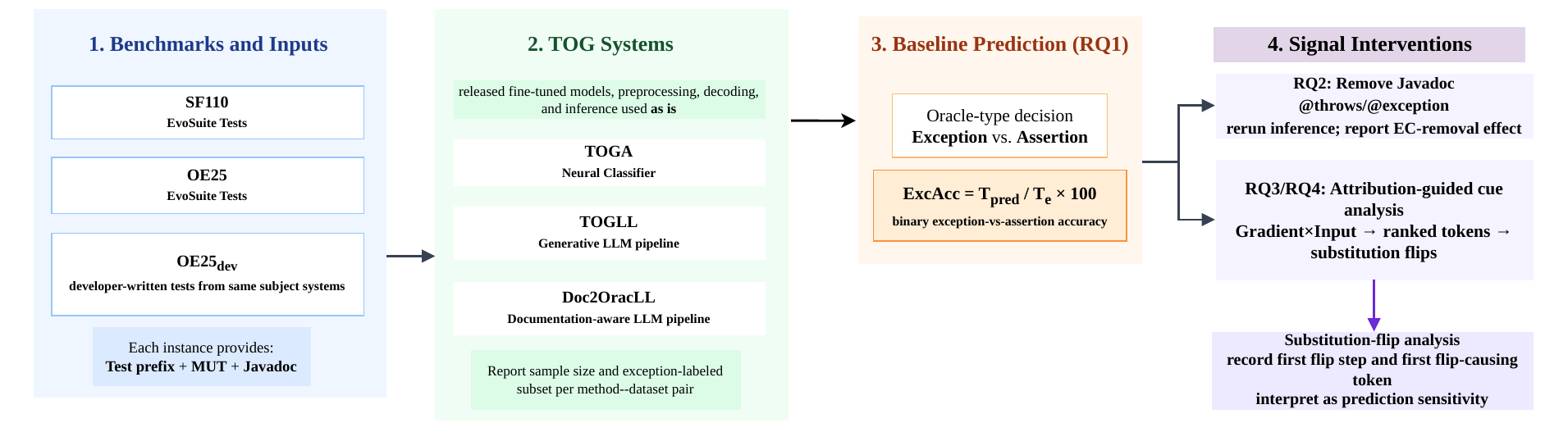}
    \caption{Overview of our approach.} 
    \label{fig:approach}
\end{figure*}

\section{Approach}
\label{sec:approach}

Our goal is to identify which input signals drive exception-oracle prediction in neural and LLM-based TOG. In particular, we ask whether structured exception documentation---Javadoc \texttt{@throws} clauses---provides the main predictive signal, or whether models rely more on recurring input cues. Figure~\ref{fig:approach} summarizes our pipeline; in this section, we discuss the benchmarks, TOG systems under study, signal analysis, and evaluation metrics.

\subsection{Benchmarks}
\label{benchmark}

We evaluate three Java benchmarks. Each benchmark sample is represented as
\begin{equation}
s_i = (x_i, y_i, m_i), \qquad x_i = \langle p_i, \mu_i, d_i \rangle .
\end{equation}

\noindent Here, $x_i$ is the model input, $y_i$ is the ground-truth oracle-type label, and $m_i$ stores metadata. The input $x_i$ contains three components: the test prefix $p_i$, the method under test (MUT) $\mu_i$, and the associated Javadoc $d_i$. The test prefix is the test code retained after removing the oracle block; it includes setup statements and the call to the method under test $\mu_i$ whose oracle type must be predicted.

\paragraph{SF110~\cite{TOGLL-artifact-2025,togll-cse-25,Doc2OracLL-fse-25}.}
SF110 consists of 110 real-world Java systems and was originally curated to evaluate EvoSuite. It has since been adopted in LLM-based TOG studies, including \textsc{TOGLL} and \textsc{Doc2OracLL}. We include SF110 to maintain comparability with these prior evaluations. SF110 contains 159,073 input samples; its test prefixes are generated by EvoSuite, while the focal code and documentation come from the original developer-written projects.

\paragraph{OE25~\cite{hossain2023neural,Hossain2023_replication,TOGLL-artifact-2025,togll-cse-25}.}
OE25 is a curated benchmark from 25 real-world Java systems: 17 Apache Commons projects and 8 additional GitHub projects, covering 56 Maven modules. Similar to SF110, its test prefixes are generated by EvoSuite, while the focal code and documentation are developer-written. The benchmark contains 223,557 samples and has been used by prior test-oracle generation studies for unseen inference~\cite{togll-cse-25,konstantinou2024llms}. We include OE25 because it provides a larger and more documentation-rich generated-test setting than SF110, while still providing the same input components required by the evaluated TOG systems described in Section~\ref{tog-systems}.

\paragraph{OE25\textsubscript{dev} (Our Contribution).}
OE25\textsubscript{dev} is curated from developer-written tests in the same 25 systems as OE25: 17 Apache Commons projects and 8 GitHub projects across 56 Maven modules. Unlike SF110 and OE25, whose test prefixes are EvoSuite-generated, OE25\textsubscript{dev} is mined from upstream project test suites authored and maintained by the original project contributors. EvoSuite-generated tests mostly follow consistent templates, and prior evaluations of \textsc{TOGA}, \textsc{TOGLL}, and \textsc{Doc2OracLL} have largely relied on such generated tests~\cite{Dinella2022TOGA, togll-cse-25, Doc2OracLL-fse-25}. OE25\textsubscript{dev} lets us test whether signal patterns observed on generated-test benchmarks persist in developer-written tests that are more diverse and realistic.

We identify 100,118 developer-written test methods. For each test, we remove the oracle, preserve the developer-written prefix, and recover the MUT and corresponding Javadoc. This work uses the \emph{Single} variant because the evaluated TOG systems assume one test prefix and a single oracle (assertion or exception) per instance. The Single variant contains 77,299 dataset instances across six oracle categories: \texttt{ASSERTION\_ONLY} (73,023), \texttt{MUST\_THROW} (2,899), \texttt{FAIL\_ONLY} (107), \texttt{MUST\_NOT\_THROW} (694), \texttt{IF\_THROWN\_ASSERT} (257), and \texttt{MUST\_THROW\_WITH\_PROPERTIES} (319). These categories capture assertion-only tests, expected-exception tests, no-exception tests, assertions over caught exceptions, expected exceptions with property checks, and fail-based oracle patterns. Our analysis focuses on \texttt{MUST\_THROW} because it maps directly to the binary exception-vs.-assertion decision supported by all three systems. The remaining categories, together with the Multi, Mixed, and Custom variants, are released for future work on richer oracle-generation settings that the evaluated systems do not uniformly support.

\subsection{TOG Systems Under Study}
\label{tog-systems}

We evaluate three representative TOG systems: \textsc{TOGA}~\cite{Dinella2022TOGA}, a neural classifier-based approach; \textsc{TOGLL}~\cite{togll-cse-25}, a generative LLM-based pipeline that uses the test prefix, focal method, and documentation; and \textsc{Doc2OracLL}~\cite{Doc2OracLL-fse-25}, a documentation-aware generative LLM pipeline. We select these systems because they are peer-reviewed Java TOG pipelines, cover both classifier-based and generative paradigms, operate over the input components needed for our study, and allow us to compare signal reliance across different model families and sizes.

We use the released fine-tuned models, preprocessing steps, decoding settings, and inference protocols of each system \textit{as is}---without modifying their original configurations---to preserve comparability with prior work~\cite{toga-replication-package, TOGLL-artifact-2025}. Because the model sizes, token budgets, preprocessing steps, and released evaluation subsets vary, the number of samples processed by each method can differ across method--dataset pairs. We therefore report the sample size and exception-labeled subset for each method--dataset pair, and interpret comparisons within those reported evaluation sets. We also repeat each configuration three times to check run-to-run stability.

\subsection{Signal Analysis}
\label{counterfactual-interventions}

We analyze two signal groups: Javadoc exceptional-behavior clauses (ECs), namely \texttt{@throws} and \texttt{@exception} clauses, and recurring non-EC cues in the test prefix, MUT, and Javadoc. EC removal is applied to all three systems, while attribution-guided cue analysis is applied to the generative systems in RQ3 and RQ4 under their respective evaluation settings. To estimate the role of structured exception documentation, we remove all ECs from the Javadoc, keep the rest of the input unchanged, rerun inference, and measure the change in exception-oracle accuracy.

To identify other cues, we apply Gradient$\times$Input attribution~\cite{shrikumar2017learning} to correctly predicted exception-oracle instances. We rank tokens by absolute attribution score and substitute ranked tokens one at a time until the prediction changes from \textsc{Exception} to a non-exception prediction, or until the substitution budget is reached. For each flipped instance, we also record the first flip-causing token and classify it as a structural token, a predefined semantic token, or an uncategorized lexical token. This measures prediction sensitivity and cue concentration, and helps identify whether correct predictions depend on meaningful evidence or recurring input cues.

\subsection{Evaluation Metrics}
\label{sec:metrics}

We report three metrics: exception-oracle accuracy, EC-removal effect, and substitution-flip sensitivity.

\emph{Exception-oracle accuracy} measures the first-stage oracle-type prediction problem. It is computed only over instances whose ground truth requires an exception oracle. If $T_e$ is the number of exception-labeled instances and $T_{pred}$ is the number predicted as requiring an exception oracle, then:

\begin{equation}
\textsc{ExcAcc}=\frac{T_{pred}}{T_e}\times 100.
\label{eq:exception_accuracy}
\end{equation}

A prediction is counted as correct when an exception-labeled instance is classified as \textsc{Exception}, not \textsc{Assertion}. This metric does not measure semantic equivalence, compilability, oracle style, or exact exception-type matching.

\emph{EC-removal effect} measures how much exception-oracle accuracy changes after removing Javadoc \texttt{@throws}/\texttt{@exception} clauses while keeping the rest of the input unchanged. We report this change in percentage points. We use $\Delta_{\mu}$ for the change over all exception-labeled samples, and $\Delta_{EC}$ for the change over the subset of exception-labeled samples that originally contained ECs. Negative values indicate that removing ECs reduces exception-oracle accuracy.

\emph{Substitution-flip sensitivity} measures how many high-attribution tokens must be substituted before a correct exception prediction changes. For each correctly predicted exception instance, we substitute ranked tokens until the prediction changes from \textsc{Exception} to a non-exception prediction, or until the substitution budget is reached. We report the percentage of predictions that flip, the mean and median flip step, and the normalized percentage of eligible tokens substituted before a flip. Lower flip steps or lower normalized percentages indicate more concentrated reliance on a small number of cues.

For each flipped instance, we also record the first flip-causing token and classify it as structural, predefined semantic, or uncategorized lexical. This cue-type distribution helps distinguish predictions driven by meaningful exception-related evidence from predictions driven by recurring structural or lexical patterns.

\section{Experimental Study}
\label{experimental-study}

We empirically study exception-oracle prediction along four dimensions: baseline exception-oracle accuracy (RQ1), the contribution of Javadoc exceptional-behavior clauses (ECs), namely \texttt{@throws} clauses (RQ2), non-EC input cues that drive predictions (RQ3), and the generalizability of these cues across evaluated model families, sizes, and datasets (RQ4).

\subsection{RQ1: Baseline exception-oracle accuracy.}
\label{sec:rq1}

We first establish the baseline accuracy of exception-oracle prediction. This step is necessary before analyzing which contextual signals drive exception-oracle decisions: we must first determine how accurately the released systems make the binary exception-vs.-assertion decision. We evaluate three representative TOG systems, \textsc{TOGA}, \textsc{TOGLL}, and \textsc{Doc2OracLL}, on SF110, OE25, and OE25\textsubscript{dev}.

\begin{table*}[t]
\centering
\caption{\mrnew{Exception-oracle prediction accuracy across TOG systems and datasets. Accuracy is computed only over exception-labeled instances and measures the binary exception-vs.-assertion decision. Sample counts report the method-specific evaluable subset. Run Accuracy reports accuracy for each of three executions; Mean/Var./SD report the mean, variance, and standard deviation of those executions.}}
\label{tab:rq1}
\small
\setlength{\tabcolsep}{5pt}
\renewcommand{\arraystretch}{1.25}
\begin{tabular}{@{}ccccccc@{}}
\toprule
\textbf{Dataset} &
\textbf{TOG System} &
\thead{\textbf{Evaluation}\\\textbf{Setting}} &
\thead{\textbf{Evaluable}\\\textbf{Samples}} &
\thead{\textbf{Exception}\\\textbf{Samples}} &
\thead{\textbf{Run Accuracy}\\\textbf{(\%)}} &
\thead{\textbf{Mean / Var. / SD}\\\textbf{(\%)}} \\
\midrule
\multirow{3}{*}{\rotatebox[origin=c]{90}{\textbf{SF110}}}

& \modelcell{togablue}{TOGA}{CodeBERT}
& Unseen & 159{,}073 & 39{,}606  & 11.81 / 11.81 / 11.81 & 11.81 / 0 / 0 \\

\cmidrule(lr){2-7}
& \modelcell{togllgreen}{TOGLL}{CodeParrot-110M}
& Seen & 7{,}767 & 1{,}902 & 100 / 100 / 100 & 100 / 0 / 0 \\
\cmidrule(lr){2-7}
& \modelcell{docpurple}{Doc2OracLL}{CodeLlama-7B}
& Seen & 2{,}810 & 721 & 100 / 100 / 100 & 100 / 0 / 0 \\
\midrule
\multirow{3}{*}{\rotatebox[origin=c]{90}{\textbf{OE25}}}
& \modelcell{togablue}{TOGA}{CodeBERT}
& Unseen & 223{,}557 & 21{,}082 & 18.66 / 18.66 / 18.66 & 18.66 / 0 / 0 \\
\cmidrule(lr){2-7}
& \modelcell{togllgreen}{TOGLL}{CodeParrot-110M}
& Unseen & 223{,}557 & 19{,}533 & 99.98 / 99.98 / 99.98 & 99.98 / 0 / 0 \\
\cmidrule(lr){2-7}
& \modelcell{docpurple}{Doc2OracLL}{CodeLlama-7B}
& Unseen & 223{,}557 & 19{,}240 & 99.7 / 99.7 / 99.7 & 99.7 / 0 / 0 \\
\midrule
\multirow{3}{*}{\rotatebox[origin=c]{90}{\textbf{OE25$_{\mathit{dev}}$}}}
& \modelcell{togablue}{TOGA}{CodeBERT}
& Unseen & 77{,}299 & 2{,}899 & 36.36 / 36.36 / 36.36 & 36.36 / 0 / 0 \\
\cmidrule(lr){2-7}
& \modelcell{togllgreen}{TOGLL}{CodeParrot-110M}
& Unseen & 77{,}299 & 2{,}722 & 100 / 100 / 100 & 100 / 0 / 0 \\
\cmidrule(lr){2-7}
& \modelcell{docpurple}{Doc2OracLL}{CodeLlama-7B}
& Unseen & 77{,}299 & 2{,}673 & 91.55 / 91.55 / 91.55 & 91.55 / 0 / 0 \\
\bottomrule
\end{tabular}
\end{table*}

\mrnew{
\subsubsection{Experimental Setup.}
We collected the released replication artifacts for \textsc{TOGA}~\cite{Dinella2022TOGA,toga-replication-package}, \textsc{TOGLL}~\cite{togll-cse-25,TOGLL-artifact-2025}, and \textsc{Doc2OracLL}~\cite{Doc2OracLL-fse-25}, and executed each method using its original evaluation protocol. \textsc{TOGA} is evaluated using its released CodeBERT-based pipeline. \textsc{TOGLL} uses the CodeParrot-110M configuration over the test prefix, method under test (MUT), and documentation. \textsc{Doc2OracLL} uses the CodeLlama-7B configuration over the test prefix, MUT, and documentation.

Each method--dataset combination is run \textit{three} times, yielding \(9 \times 3 = 27\) executions. These repeated runs test run-to-run stability under a fixed pipeline. Since the systems use fixed fine-tuned checkpoints, preprocessing, decoding settings, and inference protocols, we expect little or no variation under the same execution setup. Table~\ref{tab:rq1} confirms this expectation: all repeated executions within each method--dataset combination produce identical accuracies, yielding zero variance and zero standard deviation.

This deterministic behavior should be interpreted as stability under our execution setup, not as a guarantee across all environments. Changes in hardware, available GPU memory, batching, precision settings, context truncation, or inference configuration can change the processed inputs or final predictions. Therefore, the zero variance in Table~\ref{tab:rq1} indicates reproducibility under our fixed setup.

The SF110 sample counts differ across rows because the seen/unseen setting is method-relative rather than dataset-global. \textsc{TOGLL} and \textsc{Doc2OracLL} use the SF110 subsets exposed by their released seen-setting artifacts after method-specific preprocessing and filtering. \textsc{TOGA} is evaluated over the larger SF110 candidate pool exposed by its released unseen-setting pipeline. Thus, the different SF110 totals reflect different evaluable subsets produced by the released systems, not conflicting definitions of SF110.

\subsubsection{Results and Analysis.}

Table~\ref{tab:rq1} shows that exception-oracle accuracy is highly method-dependent. On SF110, \textsc{TOGA} achieves only 11.81\% accuracy in the unseen setting. In contrast, \textsc{TOGLL} and \textsc{Doc2OracLL} are evaluated in the seen setting and both achieve 100\% accuracy. Because these SF110 rows use different evaluation settings and evaluable subsets, this comparison should be read descriptively rather than as a controlled head-to-head comparison.

On OE25, both generative systems remain near-perfect in the unseen setting: \textsc{TOGLL} achieves 99.98\% accuracy and \textsc{Doc2OracLL} achieves 99.7\%, while \textsc{TOGA} reaches only 18.66\%. On OE25\textsubscript{dev}, which contains developer-written tests, \textsc{TOGLL} again achieves 100\%, \textsc{Doc2OracLL} drops to 91.55\%, and \textsc{TOGA} improves to 36.36\%. \textsc{TOGA} performs better on OE25\textsubscript{dev}, possibly because it was fine-tuned on developer-written tests and this benchmark is closer to its training distribution; however, it still lags far behind the generative systems. \textsc{Doc2OracLL} performs worse on OE25\textsubscript{dev}, possibly because it was fine-tuned on EvoSuite-generated tests, while OE25\textsubscript{dev} contains developer-written tests with more diverse and less predictable structures.

We ran each method--dataset pair three times, computed exception-oracle accuracy for each run, and then reported the mean, variance, and standard deviation across the three runs. All repeated runs produced identical accuracies, indicating that the released fine-tuned TOG pipelines are deterministic under our fixed inference setup.

Overall, the generative systems achieve very high exception-oracle accuracy across both generated and developer-written benchmarks, while the released \textsc{TOGA} pipeline performs much worse on the unseen evaluations. Model size alone does not explain the gap: CodeParrot-110M is comparable in scale to CodeBERT, yet \textsc{TOGLL} performs much better. The difference is more likely due to architecture, training objective, input formulation, and generation-based prediction. These results motivate the next research questions: \textit{if generative models can make exception-oracle decisions with near-perfect accuracy, we must determine which contextual signals they rely on to do so}.}
\begin{tcolorbox}
\textbf{RQ1 Finding:}
Exception-oracle prediction accuracy is stable under our fixed inference setup and highly method-dependent. For each method--dataset pair, the three repeated executions produce identical accuracies. \textsc{TOGLL} achieves near-perfect accuracy across all datasets: 100\% on SF110, 99.98\% on OE25, and 100\% on OE25\textsubscript{dev}. \textsc{Doc2OracLL} also performs strongly, with 100\% on SF110, 99.70\% on OE25, and 91.55\% on OE25\textsubscript{dev}. In contrast, \textsc{TOGA} achieves only 11.81\%, 18.66\%, and 36.36\% on the same datasets. This gap motivates our subsequent analysis of the signals used to make exception-oracle predictions.
\end{tcolorbox}

\begin{table*}[h]
\centering
\caption{\mrnew{Effect of removing Javadoc exceptional-behavior clauses (ECs) on exception-oracle prediction. Run Acc. reports the accuracy of each inference after EC removal. Mean/Var./SD are computed over the three runs. $\Delta_{\mu}$ is the percentage-point difference between the no-EC mean and the RQ1 baseline mean; $\Delta_{EC}$ is measured only on samples that originally contain ECs.}}
\label{tab:rq2}
\footnotesize
\setlength{\tabcolsep}{5pt}
\renewcommand{\arraystretch}{1.24}
\begin{tabular}{@{}ccccccccc@{}}
\toprule
\textbf{Dataset} &
\textbf{TOG System} &
\thead{\textbf{Exception}\\\textbf{Samples}} &
\thead{\textbf{Samples}\\\textbf{w/ EC}} &
\thead{\textbf{Run Acc.}\\\textbf{w/o EC (\%)}} &
\thead{\textbf{Mean / Var. / SD}\\\textbf{(\%)}} &
\thead{\textbf{Baseline}\\\textbf{(\%)}} &
\thead{\boldmath$\Delta_{\mu}$\\\textbf{(pp)}} &
\thead{\boldmath$\Delta_{EC}$\\\textbf{(pp)}} \\
\midrule
\multirow{3}{*}{\rotatebox[origin=c]{90}{\textbf{SF110}}}
& \rqmodelcell{rqblue}{TOGA}{CodeBERT}
& 39{,}591 & 6{,}631 & 11.81 / 11.81 / 11.81 & 11.81 / 0 / 0 & 11.81 & 0 & 0 \\
\cmidrule(lr){2-9}
& \rqmodelcell{rqgreen}{TOGLL}{CodeParrot-110M}
& 1{,}902 & 339 & 100 / 100 / 100 & 100 / 0 / 0 & 100 & 0 & 0 \\
\cmidrule(lr){2-9}
& \rqmodelcell{rqpurple}{Doc2OracLL}{CodeLlama-7B}
& 721 & 338 & 100 / 100 / 100 & 100 / 0 / 0 & 100 & 0 & 0 \\
\midrule
\multirow{3}{*}{\rotatebox[origin=c]{90}{\textbf{OE25}}}
& \rqmodelcell{rqblue}{TOGA}{CodeBERT}
& 21{,}082 & 6{,}361 & 18.65 / 18.65 / 18.65 & 18.65 / 0 / 0 & 18.66 & -0.01 & -0.03 \\
\cmidrule(lr){2-9}
& \rqmodelcell{rqgreen}{TOGLL}{CodeParrot-110M}
& 19{,}533 & 5{,}781 & 99.91 / 99.91 / 99.91 & 99.91 / 0 / 0 & 100.0 & -0.09 & -0.29 \\
\cmidrule(lr){2-9}
& \rqmodelcell{rqpurple}{Doc2OracLL}{CodeLlama-7B}
& 19{,}231 & 5{,}655 & 99.59 / 99.59 / 99.59 & 99.59 / 0 / 0 & 99.75 & -0.16 & -0.54 \\
\midrule
\multirow{3}{*}{\rotatebox[origin=c]{90}{\textbf{OE25$_{\mathit{dev}}$}}}
& \rqmodelcell{rqblue}{TOGA}{CodeBERT}
& 2{,}899 & 1{,}702 & 36.36 / 36.36 / 36.36 & 36.36 / 0 / 0 & 36.36 & 0 & 0 \\
\cmidrule(lr){2-9}
& \rqmodelcell{rqgreen}{TOGLL}{CodeParrot-110M}
& 2{,}722 & 1{,}556 & 100.0 / 100.0 / 100.0 & 100.0 / 0 / 0 & 100.0 & 0.00 & 0 \\
\cmidrule(lr){2-9}
& \rqmodelcell{rqpurple}{Doc2OracLL}{CodeLlama-7B}
& 2{,}673 & 1{,}507 & 91.55 / 91.55 / 91.55 & 91.55 / 0 / 0 & 91.62 & -0.07 & -0.23 \\
\bottomrule
\end{tabular}
\end{table*}
\subsection{RQ2: Contribution of Javadoc \texttt{@throws} clauses}
\label{sec:rq2}

RQ1 showed that exception-oracle accuracy varies widely across TOG methods and benchmarks, from 11.81\% to 100\%. Since Javadoc \texttt{@throws} clauses explicitly describe exceptional behavior as shown in Listings~\ref{lst:javadoc} and ~\ref{lst:exception-oracles}, we investigate whether these clauses explain the strong exception-oracle accuracy observed in some settings.

\subsubsection{Experimental Setup.}

For each exception-oracle sample, we construct a counterfactual input by removing all Javadoc exceptional-behavior clauses (ECs), i.e., \texttt{@throws}/\texttt{@exception} clauses, while keeping the test prefix, focal method, and remaining documentation unchanged. We then re-run inference and compare the resulting accuracy with the RQ1 baseline. Table~\ref{tab:rq2} reports the number of exception samples, the number of samples that originally contained ECs, the accuracy after EC removal, the baseline accuracy, and two differences: $\Delta_{\mu}$ over all exception samples and $\Delta_{EC}$ only over samples that originally contained ECs. \mrnew{As in RQ1, each configuration is run three times, and all runs produce identical accuracies.}

\subsubsection{Results and Analysis.}

Table~\ref{tab:rq2} shows that ECs appear in only a subset of exception samples. In SF110, EC coverage ranges from 16.75\% for \textsc{TOGA} to 46.88\% for \textsc{Doc2OracLL}. The higher percentage for \textsc{Doc2OracLL} is expected because its evaluated subset is documentation-rich, making ECs more likely to appear. In OE25, EC coverage is approximately 30\% for all three systems, ranging from 29.4\% for \textsc{Doc2OracLL} to 30.17\% for \textsc{TOGA}. In OE25\textsubscript{dev}, EC coverage is higher, ranging from 56.38\% for \textsc{Doc2OracLL} to 58.71\% for \textsc{TOGA}.

Removing ECs has little effect on accuracy. On SF110, all three systems remain unchanged. On OE25, \textsc{TOGA}, \textsc{TOGLL}, and \textsc{Doc2OracLL} change by only -0.01, -0.09, and -0.16 percentage points, respectively. On OE25\textsubscript{dev}, \textsc{TOGA} and \textsc{TOGLL} are unchanged, while \textsc{Doc2OracLL} drops by only -0.07 percentage points. Across all configurations, the largest observed $\Delta_{EC}$ is only -0.54 percentage points for \textsc{Doc2OracLL} on OE25.

Overall, removing Javadoc exceptional-behavior clauses (ECs) produces no change or only very small drops in exception-oracle accuracy. This suggests that ECs are not the primary driver of model predictions. Instead, the models appear to rely more heavily on other signals in the test prefix, focal method, and remaining documentation. \mrnew{Similar to RQ1, each method--dataset pair was run three times; we also report the mean, variance, and standard deviation.}

\begin{tcolorbox}

\textbf{RQ2 Finding:}

Javadoc exceptional-behavior clauses provide only a weak auxiliary signal for exception-oracle prediction. Removing ECs causes no change or only small drops in accuracy, with the largest observed $\Delta_{\mu}$ being -0.16 percentage points and the largest observed $\Delta_{EC}$ being -0.54 percentage points for \textsc{Doc2OracLL} on OE25.

\end{tcolorbox}

\subsection{RQ3: What Drives Exception-Oracle Predictions?}
\label{sec:rq3}

\mrnew{RQ2 shows that removing Javadoc \texttt{@throws} clauses has little effect on exception-oracle prediction. However, that intervention only reveals what the model does \emph{not} rely on. It does not identify what actually drives the prediction. RQ3 therefore asks which parts of the input are necessary for a correct exception prediction and whether those cues represent meaningful evidence of exceptional behavior or incidental artifacts of the input representation. This distinction matters because a model can achieve high benchmark accuracy while relying on fragile cues that may disappear under small changes in formatting, documentation style, or test construction.}

\subsubsection{Experimental Setup.}

\mrnew{We answer RQ3 with the five-stage pipeline as shown in Figure~\ref{fig:rq3_pipeline}. We focus on TOGLL on OE25 because RQ1 shows that this model achieves near-perfect exception-oracle accuracy, while RQ2 shows that this accuracy is not explained by Javadoc \texttt{@throws} clauses. From the 20{,}841 OE25 exception-oracle candidates, we exclude 985 samples that exceed the model's position-embedding limit. Among the remaining evaluable samples, TOGLL predicts 19{,}840 as exception oracles. The substitution analysis below uses the 19{,}805 samples that remain baseline-correct under the proxy prediction check at both analyzed layers.

\begin{figure}[h]
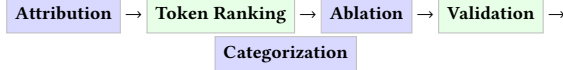

\centering
\scriptsize
\stepbox{blue!15}{\textbf{Attribution}}
$\rightarrow$
\stepbox{green!10}{\textbf{Token Ranking}}
$\rightarrow$
\stepbox{blue!15}{\textbf{Ablation}}
$\rightarrow$
\stepbox{green!10}{\textbf{Validation}}
$\rightarrow$
\stepbox{blue!15}{\textbf{Categorization}}
\caption{RQ3 analysis pipeline.}
\label{fig:rq3_pipeline}
\end{figure}

\textbf{Attribution.}
For each sample, we compute Gradient~$\times$~Input attribution in a single backward pass, scoring each token's contribution to the model's confidence in the correct \textsc{Exception} prediction. We compute attribution at the input-embedding layer and at layer~6. The input layer captures surface-level token influence, while layer~6 provides an intermediate checkpoint closer to the model's decision process.

\textbf{Token ranking.}
At each layer, we rank tokens independently by the absolute value of their attribution score. We exclude the final five token positions from candidacy because autoregressive models tend to assign high influence to tokens near the generation point regardless of their content. This prevents the ablation from measuring a trivial boundary-position artifact.

\textbf{Substitution ablation.}
We then test whether predictions are sensitive to the ranked tokens. Starting from the highest-ranked token, we substitute one additional ranked token at each step with a generic placeholder and check whether the prediction changes from \textsc{Exception} to a non-exception prediction. We record the first substitution step that flips the prediction. If no flip occurs after 30 substitutions, the sample is marked as never flipped.

\textbf{Validation.}
The substitution loop uses a fast single-token prediction check rather than full decoding. To validate this proxy, we compare it against the same beam-search decoding used in RQ1 and RQ2. We validate every detected accuracy flip and a random sample of 200 never-flipped cases from each layer.

\textbf{Categorization.}
Finally, for every flipped sample, we identify the token whose substitution first caused the prediction to flip. We classify this flip-causing token as structural, predefined semantic, or uncategorized lexical. Structural tokens include whitespace, punctuation, braces, and comment markers; predefined semantic tokens include exception-related, invalid-value, documentation, and domain-descriptive language. This yields intervention-based evidence about what the model's correct prediction truly depends on.
}
\subsubsection{Results and Analysis.}
\mrnew{
\paragraph{Correct predictions are fragile to a small number of substitutions.}
Table~\ref{tab:rq3_necessity_curve} shows that most correct exception predictions can be flipped by substituting only a few high-attribution tokens. At layer~6, 19{,}190 of 19{,}805 baseline-correct samples flip within the 30-token budget, corresponding to 96.89\%. The median flip step is only 4 tokens, and the mean is 6.04. The input-embedding layer is less concentrated: 77.40\% of samples flip within 30 substitutions, with a median of 11 and a mean of 12.37.

This difference suggests that layer~6 provides a more concentrated view of the model's exception decision. By that point, the prediction is concentrated in a smaller set of influential tokens. The median distance-from-end of the flip token is also large at both layers, 145 tokens for layer~6 and 147 for the input-embedding layer. Thus, the flips are not caused merely by tokens adjacent to the generation point; they arise from influential tokens distributed earlier in the input.}

\begin{table}[h]
\centering
\footnotesize
\caption{Necessity curve: how many top-ranked tokens must be substituted before a correct prediction flips, per layer. ``Never flipped'' samples exhausted the full 30-token budget in every case (min=max=30).}
\label{tab:rq3_necessity_curve}
\begin{tabular}{@{}lrr@{}}
\toprule
\textbf{Statistic} & \textbf{Layer 6} & \textbf{Input embed.} \\
\midrule
Baseline-correct (of 19{,}840)     & 19{,}805 (99.82\%) & 19{,}805 (99.82\%) \\
Flipped within 30 steps             & 19{,}190 (96.89\%) & 15{,}330 (77.40\%) \\
Never flipped                       & 615 (3.11\%)       & 4{,}475 (22.60\%)  \\
Flip step -- mean                   & 6.04               & 12.37              \\
Flip step -- median                 & 4                  & 11                 \\
Flip step -- std.\ dev.             & 5.24               & 7.40               \\
Flip step -- min / max              & 1 / 30             & 1 / 30             \\
Median distance-from-end of flip token & 145             & 147                \\
\bottomrule
\end{tabular}
\end{table}

\mrnew{\paragraph{The fast flip proxy closely matches full beam-search decoding.}
Table~\ref{tab:rq3_confirmation} validates the substitution proxy against full beam-search decoding. For detected flips, agreement is nearly exact: 99.98\% at layer~6 and 99.97\% at the input-embedding layer. Agreement is lower for never-flipped cases, 91.50\% at layer~6 and 96.00\% at the input layer, but still high. This means the proxy is reliable for identifying tokens whose substitution causes an accuracy flip, while the never-flipped category is slightly conservative.

This validation is important because the rest of RQ3 depends on the identity of the flip-causing token. The high agreement for flipped cases supports using the proxy to categorize the signals that the model actually depends on.}

\begin{table}[h]
\centering
\small
\caption{Validating the single-token flip proxy against full beam-search decoding.}
\label{tab:rq3_confirmation}
\begin{tabular}{@{}llrrr@{}}
\toprule
\textbf{Layer} & \textbf{Direction} & \textbf{$n$ checked} & \textbf{$n$ agree} & \textbf{\% agree} \\
\midrule
Layer 6       & Proxy-flipped        & 19{,}190 & 19{,}187 & 99.98\% \\
Layer 6       & Proxy-never-flipped  & 200      & 183      & 91.50\% \\
Input embed.  & Proxy-flipped        & 15{,}330 & 15{,}325 & 99.97\% \\
Input embed.  & Proxy-never-flipped  & 200      & 192      & 96.00\% \\
\bottomrule
\end{tabular}
\end{table}

\mrnew{\paragraph{Flip-causing tokens are overwhelmingly structural.}
Table~\ref{tab:rq3_flip_tokens} shows that the tokens that cause prediction flips are dominated by structural artifacts rather than semantic exception evidence. At layer~6, 97.81\% of flip-causing tokens are structural. Newlines alone account for 82.42\% of all layer-6 flips, followed by tabs at 6.11\%, semicolons and commas at 4.95\%, and other punctuation at 2.40\%. The input-embedding layer shows the same pattern, although less extremely: structural tokens account for 83.26\% of flips, with newlines and tabs together accounting for 62.18\%.

The predefined semantic categories are much smaller. At layer~6, C1--C8 together account for only 0.44\% of flip-causing tokens. Even after including uncategorized semantic tokens, Table~\ref{tab:rq3_struct_vs_semantic} shows that semantic tokens account for only 2.19\% of layer-6 flips. At the input-embedding layer, semantic tokens are more visible, accounting for 16.74\%, but structural tokens still dominate. This pattern indicates that the model's correct exception predictions are far more sensitive to formatting and syntactic structure than to explicit exception-related or value-related tokens.}

\begin{table}[h]
\centering
\footnotesize
\caption{What causes a prediction to flip. Each row is the share of all flip-causing tokens that fall into that category, at each layer, over the full population (input embedding: $n=$15{,}330 flips; layer 6: $n=$19{,}190 flips). Structural rows are shaded.}
\label{tab:rq3_flip_tokens}
\begin{tabular}{@{}lrr@{}}
\toprule
\textbf{Category} & \textbf{Layer 6} & \textbf{Input embed.} \\
\midrule
\rowcolor{gray!15} Newline (\texttt{\textbackslash n})       & 82.42\% & 42.94\% \\
\rowcolor{gray!15} Tab (\texttt{\textbackslash t})            & 6.11\%  & 19.24\% \\
\rowcolor{gray!15} Semicolon / comma                         & 4.95\%  & 2.43\%  \\
\rowcolor{gray!15} Other punctuation                         & 2.40\%  & 6.21\%  \\
             Uncategorized semantic                          & 1.74\%  & 12.92\% \\
\rowcolor{gray!15} Brace (\texttt{\{}\,\texttt{\}})           & 1.53\%  & 10.62\% \\
\rowcolor{gray!15} Comment marker                            & 0.33\%  & 1.28\%  \\
             C2 -- Invalid values                            & 0.17\%  & 0.72\%  \\
             C1 -- Exception related                         & 0.13\%  & 1.87\%  \\
             C7 -- Docstring keywords                        & 0.07\%  & 0.44\%  \\
\rowcolor{gray!15} Parenthesis                               & 0.06\%  & 0.55\%  \\
             C4 -- Extract-value kw.                         & 0.05\%  & 0.46\%  \\
             C5 -- Placeholder vars.                         & 0.02\%  & 0.12\%  \\
             C8 -- Domain-descriptive                        & 0.01\%  & 0.04\%  \\
             C6 -- I/O stream                                & 0.01\%  & 0.12\%  \\
             C3 -- Value-altering                            & 0.01\%  & 0.05\%  \\
\midrule
\textit{All structural (shaded rows)}          & \textit{97.81\%} & \textit{83.26\%} \\
\textit{All predefined semantic (C1--C8)}  & \textit{0.44\%}  & \textit{4.34\%}  \\
\bottomrule
\end{tabular}
\end{table}

\begin{table}[h]
\centering
\small
\caption{Aggregate structural vs.\ semantic share of flip-causing tokens. The semantic row includes both predefined semantic categories and uncategorized semantic tokens.}
\label{tab:rq3_struct_vs_semantic}
\begin{tabular}{@{}lrr@{}}
\toprule
& \textbf{Layer 6} & \textbf{Input embed.} \\
\midrule
Structural & 97.81\% (18{,}769/19{,}190) & 83.26\% (12{,}764/15{,}330) \\
Semantic   & 2.19\% (421/19{,}190)       & 16.74\% (2{,}566/15{,}330)  \\
\bottomrule
\end{tabular}
\end{table}

\mrnew{\paragraph{The raw flip-causing tokens confirm the structural pattern.}
Table~\ref{tab:rq3_top_raw_tokens} gives the same conclusion at the individual-token level. At layer~6, the top flip-causing token is the newline token, responsible for 81.86\% of all flips. The next most common tokens are tab, close-parenthesis plus semicolon, space plus open brace, and empty-call plus semicolon. These are layout and syntax markers, not exception-specific words. At the input-embedding layer, newlines and tabs again dominate, followed by space plus open brace and close-parenthesis plus semicolon. The first clearly semantic token in the input-layer top ten is \texttt{Throwable}, and it accounts for only 1.01\% of flips.

This result sharpens the interpretation of RQ2. Removing \texttt{@throws} clauses has little effect not because the model necessarily learned robust exception semantics, but because the prediction can be supported by other recurring cues that remain in the input. Many of these cues are structural features of the generated test prefix, focal method, and documentation layout. The model's high accuracy therefore reflects sensitivity to recurring input patterns, not only to meaningful exception-triggering evidence.}

\begin{table}[h]
\centering
\footnotesize
\caption{Top 10 individual flip-causing tokens (raw, uncategorized), per layer. \textvisiblespace\ marks a leading space (BPE word-boundary marker).}
\label{tab:rq3_top_raw_tokens}
\resizebox{.99\columnwidth}{!}{%
\begin{tabular}{@{}c l l r@{\hspace{1.5em}} l l r@{}}
\toprule
& \multicolumn{2}{c}{\textbf{Layer 6}} & \textbf{\%} & \multicolumn{2}{c}{\textbf{Input embedding}} & \textbf{\%} \\
\cmidrule(lr){2-4} \cmidrule(lr){5-7}
\textbf{Rank} & \textbf{Token} & \textbf{Description} & & \textbf{Token} & \textbf{Description} & \\
\midrule
1  & \texttt{\textbackslash n}                          & newline               & 81.86 & \texttt{\textbackslash n}                          & newline               & 42.56 \\
2  & \texttt{\textbackslash t}                          & tab                   & 6.11  & \texttt{\textbackslash t}                          & tab                   & 19.24 \\
3  & \texttt{);}                                        & close-paren + semi.   & 4.65  & \texttt{\textvisiblespace\{}                       & space + open-brace    & 10.47 \\
4  & \texttt{\textvisiblespace\{}                       & space + open-brace    & 1.39  & \texttt{);}                                        & close-paren + semi.   & 2.17  \\
5  & \texttt{();}                                       & empty call + semi.    & 1.06  & \texttt{\textvisiblespace void}                    & \emph{void} keyword   & 1.87  \\
6  & \texttt{\textbackslash n\textvisiblespace\textvisiblespace\textvisiblespace} & newline + indent (3sp) & 0.43 & \texttt{();}                             & empty call + semi.    & 1.38  \\
7  & \texttt{;}                                         & semicolon             & 0.29  & \texttt{((}                                        & double open-paren     & 1.06  \\
8  & \texttt{");}                                       & close-quote + semi.   & 0.28  & \texttt{\textvisiblespace Throwable}               & \emph{Throwable} type & 1.01  \\
9  & \texttt{));}                                       & double close + semi.  & 0.27  & \texttt{()}                                        & empty call            & 0.87  \\
10 & \texttt{\textvisiblespace\textvisiblespace\textvisiblespace} & indent (3 spaces)   & 0.22  & \texttt{public}                                    & keyword               & 0.85  \\
\bottomrule
\end{tabular}%
}
\end{table}

\mrnew{\paragraph{Interpretation.}
The RQ3 results show that the strongest causal signals are not necessarily the most semantically meaningful signals. Tokens such as \texttt{Throwable}, \texttt{throw}, or invalid-value terms may provide genuine evidence of exceptional behavior, but they rarely cause flips in the layer-6 analysis. Instead, the model's predictions are most fragile to substitutions of formatting and syntactic boundary tokens. This does not mean that newline characters are themselves semantic explanations for exceptions. Rather, it suggests that TOGLL has learned to exploit regularities in how exception-oracle examples are formatted and structured.

This finding is important for evaluation. A model that uses structural regularities can appear highly accurate on a benchmark while relying on cues that may not transfer to differently formatted tests, developer-written tests, or alternative prompting formats. RQ3 therefore complements RQ2: Javadoc \texttt{@throws} clauses are not the primary driver, but the replacement signal is often not robust semantic evidence either.

\begin{tcolorbox}
\textbf{RQ3 Finding:}
TOGLL's correct exception-oracle predictions are highly sensitive to a small number of high-attribution tokens, especially at layer~6, where 96.89\% of predictions flip within 30 substitutions and the median flip step is only 4. The flip-causing tokens are overwhelmingly structural: 97.81\% at layer~6 and 83.26\% at the input-embedding layer. Newlines alone account for 82.42\% of layer-6 flips. In contrast, predefined semantic categories account for only 0.44\% of layer-6 flips. Thus, TOGLL's exception-oracle accuracy is driven more by recurring structural cues in the input than by explicit exception-related semantics.
\end{tcolorbox}}

\subsection{RQ4: Does Signal-Cue Reliance Generalize?}
\label{sec:rq4}

\begin{table*}[h]
\centering
\footnotesize
\caption{Flip concentration across models and datasets at the middle layer. Tested samples are baseline-correct under the substitution proxy.}
\label{tab:rq4_flip_concentration}
\begin{tabular}{@{}lrrrrrr@{}}
\toprule
\textbf{Model / Dataset} &
\textbf{Processed} &
\textbf{Tested} &
\textbf{Flipped (\%)} &
\textbf{Median step} &
\textbf{Median eligible subst. (\%)} &
\textbf{IQR eligible subst. (\%)} \\
\midrule
\textsc{TOGLL} / OE25 & 19{,}840 & 19{,}805 & 96.89 & 4 & 2.44 & 1.39--4.03 \\
\textsc{TOGLL} / OE25\textsubscript{dev} & 2{,}773 & 2{,}773 & 98.77 & 3 & 1.84 & 0.82--3.70 \\
\textsc{Doc2OracLL} / OE25 & 2{,}500 & 1{,}534 & 92.89 & 198 & 90.13 & 87.19--92.89 \\
\textsc{Doc2OracLL} / OE25\textsubscript{dev} & 2{,}592 & 2{,}233 & 90.64 & 199 & 92.13 & 88.64--95.37 \\
\bottomrule
\end{tabular}
\end{table*}

\mrnew{RQ3 shows that \textsc{TOGLL} on OE25 is highly sensitive to a small number of high-attribution tokens, and that the first flip-causing tokens are overwhelmingly structural rather than explicit exception-related semantics. RQ4 asks whether this reliance is specific to one model and one generated-test dataset, or whether it persists across model families and more realistic test data. We therefore repeat the RQ3 analysis for \textsc{TOGLL} and \textsc{Doc2OracLL} on both OE25 and OE25\textsubscript{dev}. This compares a smaller CodeParrot-110M model with a larger CodeLlama-7B model, and generated tests with developer-written tests.

\subsubsection{Experimental Setup.}

For each model--dataset pair, we apply the same attribution-guided pipeline from RQ3: Gradient~$\times$~Input attribution, layer ranking, substitution-flip analysis, and categorization of the first flip-causing token. Each analysis uses samples that the corresponding model predicts correctly as exception oracles and that remain baseline-correct under the fast proxy used for substitution testing.

\textbf{Layer selection.}
For \textsc{TOGLL}, we reuse layer~6 from RQ3. For \textsc{Doc2OracLL}, we select layer~16 after a layer sweep over 250 correctly predicted exception-oracle samples. Layer~16 is the midpoint of CodeLlama-7B's 32 layers and is the layer where top-attributed tokens most often fall into known semantic categories. We also analyze the input-embedding layer as a reference point, but report the middle-layer results as the main comparison.

\textbf{Substitution budget.}
\textsc{TOGLL} uses the same 30-token substitution budget as in RQ3. This budget is too small for \textsc{Doc2OracLL}: with only 30 substitutions, \textsc{Doc2OracLL} flips almost no predictions. A diagnostic sweep showed that \textsc{Doc2OracLL} predictions often require substituting a large fraction of eligible tokens before flipping. We therefore use a 500-token budget for \textsc{Doc2OracLL}, stopping earlier when no eligible tokens remain.

\textbf{Comparison metrics.}
Raw substitution counts are not directly comparable because input lengths differ across models and datasets. We therefore report both the median flip step and the percentage of each sample's eligible tokens that must be substituted before the prediction flips. Lower percentages indicate concentrated reliance on a small number of cues; higher percentages indicate a more distributed decision. We also report the type of the first flip-causing token. Structural tokens include whitespace, punctuation, braces, parentheses, statement delimiters, and comment markers. Non-structural tokens include predefined semantic categories, such as exception-related, invalid-value, documentation-related, domain-descriptive, declaration-syntax, and literal-value tokens, as well as uncategorized lexical tokens.

As in RQ3, flips are measured with the fast substitution proxy and the proxy is additionally checked with beam-search confirmation. 

\subsubsection{Results and Analysis.}

\paragraph{Cue reliance generalizes, but concentration differs sharply.}
Table~\ref{tab:rq4_flip_concentration} shows that most correct predictions can be flipped within the corresponding substitution budget for all four model--dataset pairs. Thus, cue reliance is not limited to \textsc{TOGLL} or to generated tests: both models depend on input cues.

The difference is how concentrated that reliance is. \textsc{TOGLL} remains highly concentrated across both datasets. On OE25, the median flip step is only 4 tokens, corresponding to 2.44\% of eligible tokens. On OE25\textsubscript{dev}, the median flip step is even smaller, 3 tokens, corresponding to 1.84\% of eligible tokens. Thus, the concentrated reliance observed in RQ3 is not an artifact of generated tests; it also appears on developer-written tests for the same model.

\textsc{Doc2OracLL} behaves differently. Its predictions also flip in most cases, 92.89\% on OE25 and 90.64\% on OE25\textsubscript{dev}, but the median flip step is about 200 tokens in both datasets. Normalized by input length, roughly 90\% of eligible tokens must be substituted before the prediction changes. Therefore, \textsc{Doc2OracLL} is not insensitive to input cues, but its decision is much more distributed than \textsc{TOGLL}'s.}

\begin{table}[h]
\centering
\footnotesize
\caption{Aggregate type of first flip-causing token at the middle layer. Non-structural includes predefined semantic and uncategorized lexical tokens.}
\label{tab:rq4_struct_nonstruct}
\begin{tabular}{@{}lrr@{}}
\toprule
\textbf{Model / Dataset} &
\textbf{Structural} &
\textbf{Non-structural} \\
\midrule
\textsc{TOGLL} / OE25 & 97.81\% & 2.19\% \\
\textsc{TOGLL} / OE25\textsubscript{dev} & 93.61\% & 6.39\% \\
\textsc{Doc2OracLL} / OE25 & 23.44\% & 76.56\% \\
\textsc{Doc2OracLL} / OE25\textsubscript{dev} & 23.37\% & 76.63\% \\
\bottomrule
\end{tabular}
\end{table}

\mrnew{\paragraph{The type of cue reliance changes across models.}
Table~\ref{tab:rq4_struct_nonstruct} compares the aggregate type of the first flip-causing token. \textsc{TOGLL} remains overwhelmingly structural on both datasets: 97.81\% of flip-causing tokens are structural on OE25, and 93.61\% are structural on OE25\textsubscript{dev}. The developer-written benchmark reduces the structural share slightly, but structural tokens still dominate. Thus, making the test data more realistic does not remove \textsc{TOGLL}'s structural shortcut pattern.

\textsc{Doc2OracLL} shows a different profile. Only 23.44\% of its OE25 flip-causing tokens and 23.37\% of its OE25\textsubscript{dev} flip-causing tokens are structural. Most flip-causing tokens are non-structural lexical or semantic tokens. This does not mean that \textsc{Doc2OracLL} is necessarily grounded in exception semantics: many non-structural tokens are uncategorized lexical cues rather than clearly exception-triggering evidence. Still, the failure mode differs from \textsc{TOGLL}: the larger model does not rely primarily on formatting tokens.

\begin{table*}[t]
\centering
\footnotesize
\caption{Concentration of raw flip-causing tokens at the middle layer.}
\label{tab:rq4_token_concentration}
\begin{tabular}{@{}lrrrrl@{}}
\toprule
\textbf{Model / Dataset} &
\textbf{Unique flip tokens} &
\textbf{Top-1 share} &
\textbf{Top-3 share} &
\textbf{Top token} &
\textbf{Top-3 tokens} \\
\midrule
\textsc{TOGLL} / OE25 & 272 & 81.86\% & 92.62\% & \texttt{\textbackslash n} &
\texttt{\textbackslash n}, \texttt{\textbackslash t}, \texttt{);} \\
\textsc{TOGLL} / OE25\textsubscript{dev} & 56 & 81.38\% & 91.64\% & \texttt{\textbackslash n} &
\texttt{\textbackslash n}, \texttt{\textbackslash t}, \texttt{new} \\
\textsc{Doc2OracLL} / OE25 & 422 & 4.00\% & 11.37\% & \texttt{(} &
\texttt{(}, \texttt{\textless0x0A\textgreater}, \texttt{.} \\
\textsc{Doc2OracLL} / OE25\textsubscript{dev} & 610 & 4.30\% & 10.97\% & \texttt{\textless0x0A\textgreater} &
\texttt{\textless0x0A\textgreater}, \texttt{.}, \texttt{(} \\
\bottomrule
\end{tabular}
\end{table*}

\begin{table*}[h]
\centering
\footnotesize
\caption{Spearman rank correlation ($\rho$) of cue distributions across generalization axes.}
\label{tab:rq4_correlations}
\begin{tabular}{@{}llr@{}}
\toprule
\textbf{Axis} & \textbf{Comparison} & \boldmath$\rho$ \\
\midrule
Cross-dataset, category & \textsc{TOGLL}: OE25 vs.\ OE25\textsubscript{dev} & 0.720 \\
Cross-dataset, category & \textsc{Doc2OracLL}: OE25 vs.\ OE25\textsubscript{dev} & 0.734 \\
Cross-dataset, raw token & \textsc{TOGLL}: OE25 vs.\ OE25\textsubscript{dev} & -0.054, n.s.\ ($n=299$) \\
Cross-dataset, raw token & \textsc{Doc2OracLL}: OE25 vs.\ OE25\textsubscript{dev} & -0.285, $p<0.0001$ ($n=868$) \\
Cross-model, category & OE25: \textsc{TOGLL} vs.\ \textsc{Doc2OracLL} & 0.846 \\
Cross-model, category & OE25\textsubscript{dev}: \textsc{TOGLL} vs.\ \textsc{Doc2OracLL} & 0.825 \\
Cross-layer, category & \textsc{TOGLL}/OE25: input embedding vs.\ middle layer & 0.832 \\
Cross-layer, category & \textsc{TOGLL}/OE25\textsubscript{dev}: input embedding vs.\ middle layer & 0.867 \\
Cross-layer, category & \textsc{Doc2OracLL}/OE25: input embedding vs.\ middle layer & 0.993 \\
Cross-layer, category & \textsc{Doc2OracLL}/OE25\textsubscript{dev}: input embedding vs.\ middle layer & 1.000 \\
\bottomrule
\end{tabular}
\end{table*}

\paragraph{Single-token dominance is specific to TOGLL}
Table~\ref{tab:rq4_token_concentration} explains why the two models should not be interpreted as relying on cues in the same way. For \textsc{TOGLL}, one raw token dominates the flips in both datasets. The top flip-causing token, newline, accounts for more than 81\% of all flips on both OE25 and OE25\textsubscript{dev}; the top three tokens account for more than 91\%. This is the same structural shortcut pattern found in RQ3.

For \textsc{Doc2OracLL}, no comparable single-token shortcut appears. The top flip-causing token accounts for only 4.00\% of flips on OE25 and 4.30\% on OE25\textsubscript{dev}; the top three tokens account for only about 11\%. The number of unique flip-causing tokens is also much larger: 422 on OE25 and 610 on OE25\textsubscript{dev}, compared with 272 and 56 for \textsc{TOGLL}. Thus, moving to the larger documentation-aware model reduces the extreme single-token structural dependence, but it does not remove reliance on recurring input cues.

\paragraph{Category-level patterns are stable, but raw tokens do not transfer.}
Table~\ref{tab:rq4_correlations} reports Spearman rank correlations over both category distributions and raw-token distributions. At the category level, cross-dataset correlations are moderate for both models: $\rho=0.720$ for \textsc{TOGLL} and $\rho=0.734$ for \textsc{Doc2OracLL}. This suggests that each model preserves a similar ordering of cue families across generated and developer-written tests.

However, the raw-token correlations tell a different story. Across datasets, raw-token correlations are near zero or negative: $\rho=-0.054$ for \textsc{TOGLL} and $\rho=-0.285$ for \textsc{Doc2OracLL}. Thus, what generalizes is not the exact identity of the raw tokens, but the model-specific mode of reliance: \textsc{TOGLL} remains concentrated and structural, while \textsc{Doc2OracLL} remains distributed and lexical.

The cross-model category correlations are also high, $\rho=0.846$ on OE25 and $\rho=0.825$ on OE25\textsubscript{dev}, but these correlations should not be read as evidence that the two models behave the same way. They reflect similar coarse category rankings, not similar magnitudes. Tables~\ref{tab:rq4_flip_concentration}--\ref{tab:rq4_token_concentration} show the substantive difference: \textsc{TOGLL} is dominated by a few structural tokens, while \textsc{Doc2OracLL} spreads reliance across hundreds of lexical cues.

\paragraph{Interpretation.}
RQ4 refines the RQ3 conclusion. The broad phenomenon generalizes: correct exception-oracle predictions are sensitive to recurring input cues across both generated and developer-written tests, and across both \textsc{TOGLL} and \textsc{Doc2OracLL}. However, the form of that reliance changes across models. \textsc{TOGLL} relies on a highly concentrated structural shortcut, with newlines alone causing most flips in both datasets. \textsc{Doc2OracLL} is harder to destabilize and does not rely on one dominant structural token; instead, its decision is distributed across many lexical cues.

This has two implications. First, dataset realism alone is not sufficient to eliminate shortcut reliance: \textsc{TOGLL} shows the same structural dominance on OE25\textsubscript{dev}. Second, moving to a larger documentation-aware model changes the failure mode but does not eliminate cue reliance. \textsc{Doc2OracLL} avoids the extreme newline-dominated shortcut observed in \textsc{TOGLL}, but its predictions still depend on recurring lexical patterns rather than demonstrably grounded exception semantics.

\begin{tcolorbox}
\textbf{RQ4 Finding:}
Cue reliance generalizes, but its form is model-dependent. Across OE25 and OE25\textsubscript{dev}, \textsc{TOGLL} flips after only 3--4 substitutions, with structural tokens causing 97.81\% and 93.61\% of flips. \textsc{Doc2OracLL} requires substituting about 90\% of eligible tokens, with reliance distributed across many non-structural lexical cues. Thus, developer-written tests do not eliminate shortcut reliance, and the larger model shifts the failure mode from concentrated structural dependence to distributed lexical dependence.
\end{tcolorbox}
}

\mrnew{
\subsection{Threats to Validity}

\textbf{Internal validity.}
Our results depend on the released artifacts for \textsc{TOGA}, \textsc{TOGLL}, and \textsc{Doc2OracLL}, and on our scripts for OE25\textsubscript{dev} construction, EC removal, attribution, substitution, cue categorization, and evaluation. Bugs in these scripts could affect reported accuracies, flip rates, or cue distributions. We mitigate this risk by cross-checking outputs against released artifacts and prior results, repeating each method--dataset configuration under a fixed inference setup, and releasing our scripts and derived data. Our substitution analysis uses a fast flip proxy; we validate detected flips against beam-search decoding.

\textbf{External validity.}
We study Java TOG systems on SF110, OE25, and OE25\textsubscript{dev}. These datasets include generated and developer-written tests, but may not represent other languages, frameworks, documentation practices, or projects with more complete exception specifications. We also evaluate released fine-tuned TOG systems, not proprietary API models or larger instruction-tuned models, which may rely on different signals.

\textbf{Construct validity.}
Exception-oracle accuracy measures only the first-stage \textsc{Exception}-vs.-\textsc{Assertion} decision; it does not measure semantic equivalence, compilability, oracle style, exact exception type, or bug-finding usefulness. Similarly, attribution-guided substitution measures prediction sensitivity, not semantic program behavior. Perturbed inputs are not intended to remain executable, and the first flip-causing token is a diagnostic cue rather than a complete explanation.

\textbf{Conclusion validity.}
The EC-removal effects are small, so we focus on their consistency across methods and datasets rather than on isolated changes. The substitution-flip results are larger, but depend on the attribution method, analyzed layer, substitution budget, and cue taxonomy. Further studies with other models, prompts, formatting transformations, and attribution methods would strengthen these conclusions.

\section{Related Work}

Prior work has studied test-oracle generation mainly by proposing generators or measuring end-to-end performance. Blasi et al.~\cite{goffi2016automatic} derive exceptional-behavior oracles from Javadoc, while neural and LLM-based systems such as \textsc{TOGA}, ChatAssert, \textsc{TOGLL}, and \textsc{Doc2OracLL}~\cite{Dinella2022TOGA,hayet2024chatassert,togll-cse-25,Doc2OracLL-fse-25} generate assertion and exception oracles using code, tests, and documentation. Recent studies also examine the usefulness of Javadoc or code context for TOG~\cite{Doc2OracLL-fse-25,molinelli2025llms}. However, these works do not isolate what actually drives exception-oracle prediction: whether models rely on explicit exceptional-behavior clauses such as \texttt{@throws}, or on other recurring cues in the test prefix, code, and documentation. To our knowledge, this is the first large-scale intervention-based study of the signals behind exception-oracle prediction in neural and LLM-based TOG, combining \texttt{@throws} removal, attribution and ablation, and a developer-written benchmark to explain exception-oracle performance rather than only measure it.}

\mrnew{

\section{Practical Implications}
\label{sec:implications}

Our findings suggest that future TOG systems should be judged not only by whether they predict the right oracle type, but also by \textit{whether they use the right evidence}. High exception-oracle accuracy is insufficient when models remain accurate after removing \texttt{@throws}/\texttt{@exception} clauses, the documentation signal that states when and why an exception should be thrown.

Future LLM-based TOG methods should not pass large contexts uncritically. They should control which evidence is provided, perturb non-semantic cues such as formatting and recurring lexical tokens, and verify whether the prediction follows from meaningful exception evidence rather than shortcut cues. Dataset realism and model scale alone are not enough: developer-written tests do not remove \textsc{TOGLL}'s structural shortcut, and \textsc{Doc2OracLL} changes the failure mode from single-token structural dominance to distributed lexical reliance.

Model design should be evidence-driven. A TOG system should first identify the exception-triggering precondition, relate it to the MUT signature, documentation, and feasible throw path, and then generate the oracle. Documentation should act as the specification signal; code and test-prefix cues should support that signal, not replace it.

Training and evaluation should include diverse developer-written and generated tests, hard negatives, and counterfactual cases where frequent cues such as formatting patterns, exception names, or \texttt{null} do not by themselves imply exceptions. Evaluation should also report cue-sensitivity diagnostics, such as flip rate, median flip step, eligible-token percentage, and first flip-causing token type. The goal is not only correct prediction, but prediction for the right reason.
}

\section{Conclusion}

We presented a large-scale intervention-based study of what drives exception-oracle prediction in neural and LLM-based TOG. Across \textsc{TOGA}, \textsc{TOGLL}, and \textsc{Doc2OracLL}, and across SF110, OE25, and OE25\textsubscript{dev}, removing Javadoc exceptional-behavior clauses has little effect on exception-oracle accuracy. Thus, structured exception documentation is not the primary signal behind the strong exception-oracle performance reported by current TOG systems.

Our attribution-guided substitution analysis shows what replaces that documentation signal. \textsc{TOGLL} correct predictions often flip after only a few substitutions and are dominated by structural cues such as newlines and formatting tokens, even on developer-written tests. \textsc{Doc2OracLL} is harder to flip and avoids a single dominant structural shortcut, but its predictions still depend on distributed lexical cues rather than demonstrably grounded exception semantics. These findings challenge the interpretation of high exception-oracle accuracy as robust exception understanding. Future TOG systems should therefore be designed and evaluated for evidence grounding: predictions should follow from exception-triggering conditions, method behavior, and feasible throw paths, not from incidental formatting or recurring lexical patterns.

\section{Data Availability}

To support transparency and reproducibility, we release the artifact at \url{https://doi.org/10.6084/m9.figshare.30110098}~\cite{ASE2026-artifact}. 

\balance
\bibliographystyle{ACM-Reference-Format}
\bibliography{main}

\end{document}